\documentclass[a4paper,10pt]{article}
\usepackage{latexsym}
\usepackage{amsmath}
\usepackage{amssymb}
\usepackage{amscd}

\newlength{\minitwocolumn}
\makeatletter
\@addtoreset{equation}{section}
\makeatother

\title{\bf Form factors of the $SU(2)$ invariant 
massive \\ Thirring model with boundary reflection}

\author{H. Furutsu$^*$, T. Kojima$^*$ and 
Y.-H. Quano$^\dagger$}

\date{\it ${}^*$Department of Mathematices,
College of Science and Technology, 
Nihon University, Chiyoda-ku, Tokyo 101-0062, Japan \\
\it ${}^\dagger$Department of Medical Electronics, 
Suzuka University of Medical Science \\
      \it Kishioka-cho, Suzuka 510-0293, Japan}
\begin{document}

\maketitle
\begin{abstract}
The $SU(2)$ invariant massive Thirring model 
with a boundary is considered on the basis of 
the vertex operator approach. The bosonic 
formulae are presented for the vacuum vector 
and its dual in the presence of the boundary. 
The integral representations are also given 
for form factors of the present model. 
\end{abstract}

\section{Introduction}

Integrable two dimensional field theory posesses 
an infinite set of mutually commuting integrals 
of motion. For bulk (i.e., without boundary) 
massive theories, the integrability 
is ensured by the factorized scattering condition 
or the Yang--Baxter equation, in addition to the 
unitarity and crossing symmetry condition \cite{ZZ}. 
Cherednik \cite{Che} showed that the integrability 
in the presence of reflecting boundary is ensured by 
the boundary Yang--Baxter equation (the reflection 
equation) and the Yang--Baxter equation for bulk theory. 
A systematic treatment of determining the spectrum of 
integrable models with boundary reflection was 
initiated by Sklyanin \cite{Skl} in the framework 
of the algebraic Bethe ansatz. 

The earliest studies on off-shell quantities such as 
correlation functions of integrable models with non 
trivial boundary condition involved the Ising model 
\cite{Ising-bk, B}. Correlation functions for 
an impenetrable Bose gas with Neumann/Dirichlet 
boundary conditions were obtained in \cite{K}. However, 
the success in the Ising model and the impenetrable Bose 
gas model is rather special because they are equivalent to 
the free fermion theory. 

In 1994 the method to extract off-shell 
quantities was found for other integrable models with 
a boundary. The boundary crossing symmetry condition 
was proposed in \cite{GZ} on the basis of the boundary 
bootstrap approach, in order to obtain the boundary 
vacuum vectors. Jimbo et al. \cite{JK3M} developed this 
idea to obtain correlation functions of the XXZ spin 
chain with a boundary magnetic field, using the vertex 
operator approach \cite{XXZ,JM,CORR}. 
The $U_q (\widehat{sl_n})$-generalization 
of \cite{JK3M} was given in \cite{FK}. 
The generalization to the face model was given
in \cite{MW}.
Throughout the studies on the integral formulae of 
form factors for bulk sine-Gorodon models Smirnov 
found three axioms that form factors should satisfy 
\cite{Smbk}. The boundary analogue 
of Smirnov's axioms were formulated in \cite{JK2MW}. 

Let us briefly remind you of the boundary crossing symmetry 
condition. This condition is the equation involving the 
boundary $S$-matrix ($K$-matrix) and the generators 
of the Zamolodchikov--Faddev(ZF) algebra \cite{ZZ,F}, which 
determines the boundary state, the vacuum of integrable 
models with boundary reflection. In \cite{GZ} the ZF 
generators in the condition are the asymptotic operators, 
corresponding to the type II vertex operators in the 
terminology of \cite{XXZ}. When the transfer matrix of 
boundary integrable model is expressed in terms of the 
type I vertex operators and $K$-matirx, the asymptotic 
operators in the condition should be replaced 
by the type I vertex operators in order to get integral 
formulae for correlations. Concerning the XXZ chain, 
its $U_q (\widehat{sl_n})$-generalization
and the face-model, see \cite{JK3M,FK,MW}. 

In this paper we study the $SU(2)$ invariant massive 
Thirring model with a boundary. In \cite{L} Lukaynov 
constructed two kinds of generators of the ZF algebra: 
the asymptotic operators and the local operators, the 
analogue of the type II and I vertex operators, respectively. 
He also obtained the integral formulae of form factors 
for the $SU(2)$ invariant massive Thirring model and 
the sine-Gordon model in the bulk case \cite{L}, which 
satisfy Smirnov's axioms \cite{Smbk}. In \cite{CHSWY} 
the boundary crossing symmetry condition involving the 
local operators (the analogue of the type I vertex operators) 
was solved for the $SU(2)$ invariant massive Thirring model 
to construct the boundary states in terms of free bosons. 
The physical meaning of their boundary state is thus 
unclear. In \cite{HSWY} form factors satisfying the boundary 
analogue of Smirnov's axioms \cite{JK2MW} were obtained for 
the sine-Gordon model with a boundary. 

The rest of this paper is organized as follows. 
In section 2 we review the ZF algebra for the present 
case, and give the asymptotic and the local operators 
in terms of free boson. In section 3 solving the boundary 
crossing symmetry condition involving the asymptotic 
operators (the analogue of the type II vertex operators), 
we construct boundary state and its dual of 
the $SU(2)$ invariant massive Thirring model with a 
boundary. In section 4 we present explicit integral 
formulae for form factors of the model. 

\section{Formulation of the problem}

The purpose of this section is to set up the problem, 
thereby fixing the notation. 

\subsection{The model} 

The $SU(2)$ invariant massive Thirring model with 
boundary reflection is defined by describing its 
boundary $S$-matrix ($K$-matrix) in addition to 
the bulk $S$-matrix. The bulk $S$-matrix ${S}(\beta)$ 
of the model is given by
\begin{eqnarray}
{S}(\beta)=
\frac{\Gamma\left(\frac{1}{2}+\frac{\beta}{2\pi i}\right) 
\Gamma\left(-\frac{\beta}{2\pi i}\right)}
{\Gamma\left(\frac{1}{2}-\frac{\beta}{2\pi i}\right) 
\Gamma\left(\frac{\beta}{2\pi i}\right)
}\left(
\begin{array}{cccc}
1& & & \\
&\frac{\beta}{\beta-\pi i}&-\frac{\pi i}{\beta-\pi i}&\\
&-\frac{\pi i}{\beta-\pi i}&\frac{\beta}{\beta-\pi i}&\\
& & &1
\end{array}\right).
\end{eqnarray}
The $S$-matrix satisfies
the Yang--Baxter equation: 
\begin{equation}
S_{12}(\beta_1 -\beta_2 )
S_{13}(\beta_1 -\beta_3 )
S_{23}(\beta_2 -\beta_3 )=
S_{23}(\beta_2 -\beta_3 )
S_{13}(\beta_1 -\beta_3 )
S_{12}(\beta_1 -\beta_2 ); 
\end{equation}
the unitarity symmetry: 
\begin{eqnarray}
S_{12}(\beta)S_{21}(-\beta)=1; 
\end{eqnarray}
and the crossing symmetry: 
\begin{eqnarray}
S_{12}(\pi i -\beta)=C_1\,
{}^{t_1}\!S_{12}(\beta)\,C_1,&& 
C=\left(\begin{array}{cc}0&1\\
-1&0
\end{array}\right).
\end{eqnarray}
The Zamolodchikov--Faddeev(ZF) operators 
$Z_a^*(\beta), Z^{a}(\beta)\,(a=\pm)$ 
generate the following algebra: 
\begin{equation}
\begin{array}{rcl}
Z_a^*(\beta_1)Z_b^*(\beta_2)&=&\displaystyle
\sum_{cd} S_{ab}^{cd}(\beta_1-\beta_2)
Z_d^*(\beta_2)Z_c^*(\beta_1), \\ 
Z^a(\beta_1)Z^b(\beta_2)&=&\displaystyle
\sum_{cd} Z^d(\beta_2)Z^c(\beta_1)
S^{ab}_{cd}(\beta_1-\beta_2), \\ 
Z^a(\beta_1)Z_b^*(\beta_2)&=&\displaystyle
\sum_{cd} Z_d^*(\beta_2)
S^{da}_{bc}(\beta_1-\beta_2) Z^c(\beta_1)+
g\delta^a_b \delta (\beta_1 -\beta_2 ), 
\end{array}
\end{equation}
where $g$ is a constant. 
The boundary $S$-matrix ($K$-matrix) 
${K}(\beta)$ is given by
\begin{eqnarray}
{K}(\beta)=\frac{h(-\beta)}{h(\beta)}
\left(\begin{array}{cc}
1 & \\
 & \frac{\mu-\beta}{\mu+\beta}
\end{array}\right),~~
h(\beta)=
\frac{\Gamma\left(
-\frac{\mu}{2 \pi i}-\frac{\beta}{2\pi i}\right)
\Gamma\left(\frac{1}{4}-\frac{\beta}{2\pi i}\right)
}
{\Gamma\left(\frac{1}{2}
-\frac{\mu}{2 \pi i}-\frac{\beta}{2\pi i}\right)
\Gamma\left(-\frac{\beta}{2\pi i}\right)
}.
\end{eqnarray}
The $K$-matrix satisfies the
boundary Yang--Baxter equation (reflection equation) 
\cite{Che}: 
\begin{eqnarray}
K_2(\beta_2)S_{21}(\beta_1+\beta_2)K_1(\beta_1)
S_{12}(\beta_1-\beta_2)=
S_{21}(\beta_1-\beta_2)K_1(\beta_1)
S_{12}(\beta_1+\beta_2)K_2(\beta_2),
\end{eqnarray}
the boundary unitarity symmetry: 
\begin{equation}
K(\beta)K(-\beta)=1, 
\end{equation}
and the boundary crossing symmetry \cite{GZ}: 
\begin{eqnarray}
C_{a\,-a}K_{-a}^{-a}\left(\frac{\pi i}{2}-\beta\right)=
\sum_{c=\pm}S_{c\,-c}^{a\,-a}(2\beta)
C_{-c\,c}K_c^c\left(\frac{\pi i}{2}+\beta\right).
\end{eqnarray}
{}From the same argument in \cite{GZ,JK3M} the boundary 
state $|B\rangle$ and its dual state $\langle B|$ 
should satisfy 
\begin{equation}
\begin{array}{rcl}
Z_a^* (\beta)|B\rangle&=&K_{a}^a(\beta)
Z_{a}^* (-\beta)|B\rangle,\\
\langle B|Z^{a}(-\beta)&=&
\langle B|Z^{a}(\beta)K_a^a(\beta), 
\label{eq:b-x-sym}
\end{array}
\end{equation}
where $Z_{a}^{*}(\beta)$ and $Z^a (\beta )$ is 
connected by
\begin{eqnarray}
Z^{a}(\beta)=\sum_{b=\pm} C^{ab}Z_{b}^* (\pi i+\beta).
\end{eqnarray} 

\subsection{Ultraviolet regularization} 
Following Lukyanov \cite{L} 
we consider the ultraviolet regularization 
of the original model such that the bosonic 
field has the oscillator decomposition. 

Let us fix the parameter $\epsilon>0$.
In what follows we use the abbreviations,
$x=e^{-\frac{\pi \epsilon}{2}}, 
\zeta=e^{i \epsilon \beta},
w=e^{i\epsilon \gamma}$ and $r=e^{i\epsilon \mu}$. 
We also use the symbol 
$$
(z; p_1 , \cdots , p_m )_\infty :=
\prod_{i_1 , \cdots i_m \geqq 0} 
(1-z p_1^{i_1}\cdots p_m^{i_m}), ~~~~
|p_i|<1\,\,(1\leqq i\leqq m). 
$$
The q-analog of $\Gamma$ function is defined by 
\begin{equation}
\Gamma_q(z)=(1-q)^{1-z}
\frac{(q;q)_\infty}{(q^z;q)_\infty}. 
\label{eq:q-Gamma}
\end{equation}
You can easily see that $\Gamma_q (z) \longrightarrow 
\Gamma (z)$ as $q\rightarrow 1$. 

The regularized 
bulk $S$-matrix ${S}_\epsilon(\beta)$ 
is given by
\begin{eqnarray}
{S}_\epsilon(\beta)=
\frac{1}{\sqrt{\zeta}}
\frac{g_\epsilon(-\beta)}
{g_\epsilon(\beta)}\left(
\begin{array}{cccc}
1& & & \\ &
\displaystyle
\frac{\sinh\frac{i\epsilon \beta}{2}}
{\sinh \frac{i\epsilon(i\pi-\beta)}{2}}&
\displaystyle
-\frac{\sinh \frac{\pi \epsilon}{2}}
{\sinh \frac{i\epsilon (i\pi-\beta)}{2}}&\\
&
\displaystyle
-\frac{\sinh \frac{\pi \epsilon}{2}}
{\sinh \frac{i\epsilon (i\pi-\beta)}{2}}
&
\displaystyle
\frac{\sinh\frac{i\epsilon \beta}{2}}
{\sinh \frac{i\epsilon(i\pi-\beta)}{2}}&\\
& & & 1
\end{array}\right),~~~~
g_\epsilon(\beta)=
\frac{(\zeta^{-1};x^4)_\infty}
{(x^2 \zeta^{-1};x^4)_\infty}.
\end{eqnarray}
The regularized boundary $S$-matrix ($K$-matrix) 
is given by
\begin{eqnarray}
K_\epsilon(\beta)=\frac{h_\epsilon(-\beta)}
{h_\epsilon(\beta)}
\left(\begin{array}{cc}
1 & \\
 & 
\displaystyle \frac{\zeta-r}{1-r\zeta}
\end{array}\right),~~
h_\epsilon(\beta)=
\frac{(x^2r^{-1}\zeta^{-1};x^4)_\infty 
(\zeta^{-2};x^8)_\infty}
{(r^{-1}\zeta^{-1};x^4)_\infty 
(x^2\zeta^{-2};x^8)_\infty}.
\end{eqnarray}
In the limit $\epsilon \to 0$, 
these regularized matrices behave like 
\begin{eqnarray}
{S_\epsilon}_{ab}^{cd}(\beta)\longrightarrow 
(-1)^{1+\frac{a+d}{2}}S_{ab}^{cd}(\beta),~~
K_\epsilon(\beta) \longrightarrow K(\beta).
\end{eqnarray}

\subsection{Bosonizations of the ZF algebra} 

Let us consider the following free boson \cite{L}
\begin{equation}
\phi_\epsilon (\beta )=\dfrac{1}{\sqrt{2}}
(Q-\epsilon\beta P)+\sum_{m\neq 0} 
\dfrac{2a_m \zeta^m}{i(x^{-2m}-x^{2m})}, 
\end{equation}
where the oscilator modes $a_m$ and 
zero modes $P$, $Q$ 
satisfy the commutation relations 
\begin{equation}
[a_m , a_n ]=
\dfrac{(x^m -x^{-m})(x^{2m}-x^{-2m})x^{-|m|}}{4m}
\delta_{m+n,0}, ~~~~
[P,Q]=\dfrac{1}{i}. 
\label{eq:a-com}
\end{equation}
Let $\phi_\epsilon^+ (\beta )$ and 
$\phi_\epsilon^- (\beta )$ denote the positive 
and negative frequency part of 
$\phi_\epsilon (\beta )$, respectively. Then 
we have 
\begin{equation}
[\phi_\epsilon^+ (\beta_1 ), 
\phi_\epsilon^- (\beta_2 )]=-\log g_\epsilon 
(\beta_2 -\beta_1 ). 
\end{equation}

We now introduce the elementary vertex operators 
\begin{eqnarray}
V(\beta)=\zeta^{-\frac{1}{4}}:e^{i \phi_\epsilon(\beta)}:,~~~~~
\bar{V}(\gamma)=w^{-1}:e^{-i \bar{\phi}_\epsilon(\gamma)}:, 
\end{eqnarray} 
where
\begin{equation}
\begin{array}{rcl}
\bar{\phi}_\epsilon(\gamma)
&=&\phi_\epsilon(\gamma +\frac{\pi}{2}i)+
\phi_\epsilon(\gamma -\frac{\pi}{2}i) \\
&=&\sqrt{2}(Q-\epsilon\gamma P)+
\displaystyle\sum_{m\neq 0} 
\dfrac{2a_m w^m}{i(x^{-m}-x^{m})}. 
\end{array}
\label{eq:phi-bar}
\end{equation}
The positive and negative frequency part of 
$\bar{\phi}_\epsilon (\gamma )$, 
$\bar{\phi}^+_\epsilon (\gamma )$ and 
$\bar{\phi}^-_\epsilon (\gamma )$ are also defined. 
Lukyanov \cite{L} prove that the bosonizations of 
the asymptotic operators of the ZF algebra 
(the analogue of the type II vertex operators)
are given as follows: 
\begin{equation}
\begin{array}{rcl}
Z_{\epsilon +}^*(\beta)&=&
\displaystyle\zeta^{\frac{1}{4}}V(\beta),\\
Z_{\epsilon -}^*(\beta)&=&c_\epsilon\zeta^{-\frac{1}{4}}
 \displaystyle\int_{-\frac{\pi}{\epsilon}}
^{\frac{\pi}{\epsilon}}
\frac{d\gamma}{2\pi} \left( \bar{V}(\gamma) V(\beta) 
+xV(\beta) \bar{V}(\gamma) \right) \\
&=&c_\epsilon (x-x^{-1})
\displaystyle\int_{-\frac{\pi}{\epsilon}}
^{\frac{\pi}{\epsilon}}\frac{d\gamma}{2\pi} 
\frac{\zeta^{\frac{3}{4}}}
{\left(1-w/(x \zeta)\right)\left(1-\zeta/(x w)\right)}
:\bar{V}(\gamma) V(\beta):, 
\end{array}
\label{eq:asym-gen}
\end{equation}
where $c_\epsilon$ is an irrelevant constant. 

In order to construct the local operators of the 
ZF algebra (the analogue of the type I vertex operators), 
let us also consider the fields 
\begin{equation}
\begin{array}{rcl}
\phi'_\epsilon (\alpha )&=&
-\dfrac{1}{\sqrt{2}}(Q-\epsilon\alpha P)
-\displaystyle\sum_{m\neq 0}
\dfrac{2x^{|m|}a_m }{i(x^{-2m}-x^{2m})}\xi^m , \\
\bar{\phi}'_\epsilon (\delta )&=&
\phi'_\epsilon (\delta +\frac{\pi}{2}i)+
\phi'_\epsilon (\delta -\frac{\pi}{2}i) \\
&=&-\sqrt{2}(Q-\epsilon\delta P)
-\displaystyle\sum_{m>0}
\dfrac{2x^m (a_m v^m -a_{-m}v^{-m})}
{i(x^{-m}-x^m)}, 
\end{array}
\end{equation}
where $\xi =e^{i\epsilon\alpha}$ and 
$v=e^{i\epsilon\delta}$. We now define the 
local operators $Z'_a (\alpha)$\,($a=\pm$) 
\begin{equation}
\begin{array}{rcl}
Z'_{\epsilon +} (\alpha )&=& \xi V'(\alpha), \\
Z'_{\epsilon -} (\alpha )&=& ic'_\epsilon \displaystyle\int_
{-\frac{\pi}{\epsilon}}^{\frac{\pi}{\epsilon}}
\frac{d\delta}{2\pi} v\left( 
\bar{V}' (\delta)V' (\alpha)
+V' (\alpha)\bar{V}' (\delta) \right), 
\end{array}
\label{eq:local-gen}
\end{equation}
where $c'_\epsilon$ is another irrelevant constant, and 
\begin{eqnarray}
V'(\alpha)=:e^{i\phi'_\epsilon (\alpha)}:,&&
\bar{V}'(\delta)=v^{-1}
:e^{-i \bar{\phi}'_\epsilon (\delta)}:. 
\end{eqnarray}
Those local operators satisfy the following 
commutation relations \cite{L}: 
\begin{equation}
\begin{array}{rcl}
Z'_{\epsilon a}(\alpha_1 )
Z'_{\epsilon b}(\alpha_2 )&=&
-\displaystyle\sum_{cd} S_{ab}^{cd}(\alpha_2-\alpha_1)
Z'_{\epsilon d}(\alpha_2 )
Z'_{\epsilon c}(\alpha_1 ) \\
Z^*_{\epsilon a}(\alpha )
Z'_{\epsilon b}(\beta )&=&
ab \tan \left(\frac{\pi}{4}+i\frac{\beta -\alpha}{2} 
\right) Z'_{\epsilon b}(\beta )Z^*_{\epsilon a}(\alpha ). 
\end{array}
\end{equation}
The generators (\ref{eq:asym-gen}) and 
(\ref{eq:local-gen}) act on the regularized Fock 
space ${\cal F}_\epsilon$, whose 
vacuum vector $|0\rangle$ is characterized as follows: 
$$
P|0\rangle=0,
~~a_m |0\rangle=0,~(m>0).
$$

\subsection{Observable local fields of the model}

In this paper we are interested in the soliton form
factors of the observable local field ${\cal O}$ given by
\begin{eqnarray}
G^{\cal O}_{\epsilon\,a_1\cdots a_n}
(\beta_1,\cdots,\beta_n)
=\frac{\langle B| {\cal O} Z_{\epsilon a_1}^*(\beta_1)
\cdots Z_{\epsilon a_n}^*(\beta_n)|B \rangle}
{\langle B | B\rangle}. 
\label{eq:gen-ff}
\end{eqnarray}
The space of observable local fileds are identified with 
the set of fields commuting with the type II 
vertex operators \cite{JM}, or the asymptotic operators 
of the ZF algebra \cite{L} up to a phase factor. 
In \cite{L} the set of observable local fields are 
introduced in terms of the local operators of the 
ZF algebra (the analogue of the type I vertex operators) 
as follows: 
\begin{eqnarray}
\Lambda_m(\alpha)=\frac{i}{\eta'}
\left[\begin{array}{ccc}
\frac{1}{2}& \frac{1}{2}& 1\\
\frac{a}{2}& \frac{b}{2}& m
\end{array}
\right]_{-1}
Z_{\epsilon a}'(\alpha+i\frac{\pi}{2})
Z_{\epsilon b}'(\alpha-i\frac{\pi}{2}),
\end{eqnarray}
where the Clebsch-Gordan coefficients are given by
\begin{eqnarray}
\left[\begin{array}{ccc}
\frac{1}{2}& \frac{1}{2}& 1\\
\frac{a}{2}& \frac{b}{2}& \pm1
\end{array}
\right]_{-1}&=&\delta_{a+b,\pm1},\\
\left[\begin{array}{ccc}
\frac{1}{2}& \frac{1}{2}& 1\\
\frac{a}{2}& \frac{b}{2}& 0
\end{array}
\right]_{-1}&=&(-1)^{\frac{1-a}{2}}
\frac{\delta_{a+b,0}}{\sqrt{2}}.
\end{eqnarray}
The fields $\Lambda_m(\alpha)$'s are expressed 
in terms of bosons \cite{L} 
\begin{equation}
\begin{array}{rcl}
\Lambda_1 (\alpha )&=&\xi\widetilde{V}'(\alpha), \\
\Lambda_0 (\alpha )&=&\dfrac{i}{\sqrt{2}} 
\displaystyle\int_{-\frac{\pi}{\epsilon}}
^{\frac{\pi}{\epsilon}}
\frac{d\delta}{2\pi} v\xi\left( 
\bar{V}' (\delta)\widetilde{V}' (\alpha)
-\widetilde{V}' (\alpha)\bar{V}' (\delta) \right), \\
\Lambda_{-1}(\alpha )&=&\dfrac{1}{2}
\displaystyle\int_{-\frac{\pi}{\epsilon}}
^{\frac{\pi}{\epsilon}} \frac{d\delta_1}{2\pi} 
\int_{-\frac{\pi}{\epsilon}}
^{\frac{\pi}{\epsilon}} \frac{d\delta_2}{2\pi} 
v_1 v_2 \xi \left( \bar{V}' (\delta_1 )\bar{V}' (\delta_2 )
\widetilde{V}' (\alpha) \right. \\
&&\,\,\,\left. -2\bar{V}' (\delta_1 )
\widetilde{V}' (\alpha)\bar{V}' (\delta_2) 
+\widetilde{V}' (\alpha)\bar{V}' (\delta_1 )
\bar{V}' (\delta_2 )\right), 
\end{array}
\end{equation}
where
\begin{equation}
\widetilde{V}'(\alpha)=\xi^{-1}
:e^{i\bar{\phi}'_\epsilon (\alpha)}:. 
\end{equation}
The fields $\Lambda_m$'s 
(anti-)commute with the asymptotic operators: 
\begin{equation}
\Lambda_m (\alpha )Z^*_{\epsilon a} (\beta )=(-1)^m 
Z^*_{\epsilon a} (\beta )\Lambda_m (\alpha ). 
\end{equation}

\section{Boundary states}

In this section we wish to determine the boundary
state satisfying the relation \cite{GZ,JK3M} 
\begin{eqnarray}
Z_{\epsilon\,a}^*(\beta)|B_\epsilon \rangle
&=&{K_\epsilon}_a^a(\beta)Z_{\epsilon\,a}^*(-\beta)
|B_\epsilon \rangle,~~(a=\pm).\label{def:vacuum}
\end{eqnarray}
In order to solve (\ref{def:vacuum}) we make 
the ansatz that the boundary 
state has the following form
\begin{eqnarray}
|B_\epsilon \rangle
=e^{B_\epsilon}|0\rangle,~~~
B_\epsilon=\frac{1}{2}\sum_{m=1}^\infty
\frac{\alpha_m}{[a_m,a_{-m}]}a_{-m}^2
+\sum_{m=1}^\infty
\frac{\beta_m}{[a_m,a_{-m}]}a_{-m}. 
\label{eq:|B>}
\end{eqnarray}
Note that $[a_m,a_{-m}]$ appearing in the 
denominator is a $c$-number because of 
(\ref{eq:a-com}). From the boundary Yang--Baxter 
equation, the coefficients $\alpha_m, \beta_m$
do not depend on the spectral parameter $\beta$. 
The presence of $e^{B_\epsilon}$ has 
the effect of a Bogoliubov transformation 
\begin{eqnarray}
e^{-B_\epsilon} a_n e^{B_\epsilon}=a_n+\alpha_n a_{-n}+
\beta_n,~~~
e^{-B_\epsilon} a_{-n} e^{B_\epsilon}=a_{-n}, 
\end{eqnarray}
for $n>0$. 
Using the bosonization formulae, eq. (\ref{def:vacuum}) 
for $a=+$ reduces to 
\begin{eqnarray}
e^{i \phi_+(\beta)}|B_\epsilon\rangle=
h_\epsilon(-\beta)e^{i \phi_-(-\beta)}|B_\epsilon\rangle.
\end{eqnarray}
By straightforward calculation,
the coefficients $\alpha_m, \beta_m$ are to be found
\begin{eqnarray}
\alpha_m=-1,~~\beta_m=\frac{1}{2m}(1-x^{2m})x^{-2m}r^{-m}
-\frac{1}{2m}\theta_m (1-x^m)(1-x^{2m})x^{-2m}, 
\end{eqnarray}
where 
\begin{eqnarray}
\theta_m=\left\{\begin{array}{cc}
1,& m:{\rm even},\\
0,& m:{\rm odd}.
\end{array}\right.
\end{eqnarray}
Let us prove (\ref{def:vacuum}) for $a=-$. From 
(\ref{eq:phi-bar}) and (\ref{eq:|B>}) we have 
\begin{eqnarray}
e^{-i\bar{\phi}_+(\gamma)}|B_\epsilon \rangle
=I_\epsilon(\gamma)
e^{-i\bar{\phi}_-(-\gamma)}|B_\epsilon \rangle,
\end{eqnarray}
where 
\begin{eqnarray}
I_\epsilon(\gamma)=(1-w^2)(1-w/(xr)). 
\label{eq:df-I}
\end{eqnarray}
Thus eq. (\ref{def:vacuum}) for $a=-$ reduces to 
\begin{eqnarray}
\int_{-\frac{\pi}{\epsilon}}^{\frac{\pi}{\epsilon}}
d\gamma \frac{I_\epsilon(\gamma)
(1-r\zeta)}{w(1-w/(x\zeta))
(1-1/(xw\zeta))(1-w/(x\zeta))}
e^{-i\bar{\phi}_-(\gamma)-i\bar{\phi}_-(-\gamma)}
|B_\epsilon\rangle =(\zeta \leftrightarrow \zeta^{-1}). 
\label{eq:bse-}
\end{eqnarray}
Here we change the integral variable $\gamma \to -\gamma$ 
in both sides and add it to the original one. Then 
(\ref{eq:bse-}) holds if the following the relation 
concerning the integrand 
\begin{eqnarray}
\frac{I_\epsilon(\gamma)}{I_\epsilon(-\gamma)}=
-w^2 \frac{(1-w/(xr))}{(1-1/(xrw))}. 
\label{eq:I/I}
\end{eqnarray}
Since (\ref{eq:I/I}) is compatible with (\ref{eq:df-I}), 
we have (\ref{eq:bse-}), which implies 
(\ref{def:vacuum}) for $a=-$. 

The dual boundary state $\langle B_\epsilon |$
is determined by 
\begin{eqnarray}
\langle B_{\epsilon} |Z_\epsilon^{a}(-\beta)
=
\langle B_{\epsilon} |Z_\epsilon^{a}(\beta)
K_{\epsilon\,a}^a(\beta),~~~(a=\pm). 
\label{eq:<vac|}
\end{eqnarray}
In order to solve (\ref{eq:<vac|}) we make the ansatz that 
the dual boundary state has the following form
\begin{eqnarray}
\langle B_\epsilon |=\langle 0|e^{G_\epsilon},
~~~G_\epsilon=\frac{1}{2}\sum_{m=1}^\infty
\frac{\gamma_m}{[a_m,a_{-m}]}a_{m}^2+
\sum_{m=1}^\infty \frac{\delta_m}{[a_m,a_{-m}]}a_m.
\end{eqnarray}
>From the same arguments as before,
the coefficients $\gamma_m, \delta_m$
are to be found
\begin{eqnarray}
\gamma_m=-x^{4m},~~\delta_m
=\frac{1}{2m}(1-x^{2m})x^{2m}r^{-m}+
\frac{1}{2m}\theta_m (1-x^m)(1-x^{2m}).
\end{eqnarray}
The action of the elementary operators to the dual boundary state
sre given by
\begin{eqnarray}
\langle B_\epsilon |e^{i\phi_-(i\pi-\beta)}&=&
(1-r^{-1}\zeta)h_\epsilon (-\beta)
\langle B_\epsilon |e^{i\phi_+(i\pi+\beta)},
\\
\langle B_\epsilon |e^{-i \bar{\phi}_-(i\pi+\gamma)}
&=&
I_\epsilon^*(\gamma)
\langle B_\epsilon |e^{-i \bar{\phi}_+(i\pi-\gamma)},
\end{eqnarray}	
where 
\begin{eqnarray}
I_\epsilon^*(\gamma)
=\frac{1-1/(w^{2})}{1-x/(rw)}.
\end{eqnarray}
Note that our boundary states are differnt from 
the ones constructed by Chao et al. \cite{CHSWY}, 
because they solved the condition 
\begin{equation}
Z'_{\epsilon\,a}(\beta)|B'_\epsilon \rangle
={K_\epsilon}_a^a(\beta)Z'_{\epsilon\,a}(-\beta)
|B'_\epsilon \rangle, ~~~~
\langle B'_{\epsilon} |Z'{}^{a*}_{\epsilon}(-\beta)
=\langle B'_{\epsilon} |Z'{}^{a*}_\epsilon(\beta)
K_{\epsilon\,a}^a(\beta), 
\end{equation}
instead of (\ref{eq:b-x-sym}). 
In order to compute form factors one should identify 
$|B_\epsilon \rangle$ and $\langle B_{\epsilon}|$ 
satisfying (\ref{eq:b-x-sym}) 
with the vauum and its dual of the present model. 

\section{Form Factors}

Let us consider the form factor of the form 
\begin{eqnarray}
G^{m_1\cdots m_k}_{\epsilon\,a_1\cdots a_n}
(\alpha_1 , \cdots , \alpha_k |\beta_1,\cdots,\beta_n)
=\frac{\langle B_\epsilon | \Lambda_{m_1}
(\alpha_1)\cdots 
\Lambda_{m_k}(\alpha_k) Z_{\epsilon a_1}^*(\beta_1)
\cdots Z_{\epsilon a_n}^*(\beta_n)|B_\epsilon \rangle}
{\langle B_\epsilon | B_\epsilon\rangle}, 
\label{eq:df-ff}
\end{eqnarray}
where $\Lambda_{m}(\alpha)$ ($m=0, \pm 1)$ are the 
observable local fields of the present model introduced 
in section 2.4. 

In what follows we use the following abbreviations: 
\begin{eqnarray}
\xi_j=e^{i\epsilon \alpha_j},~
v_j=e^{i\epsilon \delta_j},~
\zeta_j=e^{i\epsilon \beta_j},~
w_j=e^{i\epsilon \gamma_j}.
\end{eqnarray}
Let $A_\pm$ and $M_m$ ($m=0,\pm 1$) signify the sets 
such that 
$$
A_\pm :=\{ a_j | a_j =\pm \}, ~~~~
M_m :=\{ m_j | m_j =m \}, 
$$
and $r=\#(A_-)$, $p=2\#(M_{-1})+\#(M_0 )$. 
Then we find the integral formula for the 
form factor (\ref{eq:df-ff}) as follows: 
\begin{equation}
\begin{array}{cl}
&G^{m_1\cdots m_k}_{\epsilon\,a_1\cdots a_n}
(\alpha_1 , \cdots , \alpha_k |\beta_1,\cdots,\beta_n) \\
=&\displaystyle\prod_{a\in A_-}\int_{-\frac{\pi}{\epsilon}}
^{\frac{\pi}{\epsilon}} \frac{d\gamma_a}{2\pi} 
c_\epsilon \zeta_a^{-\frac{1}{4}}
(P_\epsilon (\zeta_a , w_a )+x) 
\prod_{j=a+1}^n P_\epsilon (\zeta_j , w_a ) \\
\times& \displaystyle\prod_{m\in M_{-1}}
\int_{-\frac{\pi}{\epsilon}}
^{\frac{\pi}{\epsilon}} \frac{d\delta^1_m}{2\pi} 
\int_{-\frac{\pi}{\epsilon}}
^{\frac{\pi}{\epsilon}} \frac{d\delta^2_m}{2\pi} 
\xi_m v_m^1 v_m^2 \left( 
P'_\epsilon (\xi_m , v_m^1) P'_\epsilon (\xi_m , v_m^2) -
2P'_\epsilon (\xi_m , v_m^1)+1 \right) \\
\times& \displaystyle\prod_{j=m+1}^k 
P'_\epsilon (\xi_j , v_m^1) P'_\epsilon (\xi_j , v_m^2) 
\prod_{m' \in M_0} \int_{-\frac{\pi}{\epsilon}}
^{\frac{\pi}{\epsilon}} \frac{d\delta_{m'}}{2\pi} 
\xi_{m'}v_{m'} \left( P'_\epsilon (\xi_{m'}, v_{m'})-1 
\right) \prod_{j=m'+1}^k P'_\epsilon (\xi_j , v_{m'}) \\
\times &R_\epsilon(\alpha_1,\cdots,\alpha_k|
\{ \delta_m^1 , \delta_m^2 \}_{m\in M_{-1}}, 
\{ \delta_{m'} \}_{m'\in M_0}|\beta_1,\cdots,\beta_n|
\{ \gamma_a\}_{a\in A_-})
\end{array}
\label{eq:int-ff}
\end{equation}
where 
\begin{equation}
P_\epsilon (\zeta , w):=
\frac{\zeta -xw}{w-x\zeta}, ~~~~
P'_\epsilon (\xi , v):=\frac{x\xi -x^{-1}v}{x^{-1}\xi-xv}, 
\end{equation}
and the auxiliary function 
\begin{eqnarray}
&&R_\epsilon(\alpha_1,\cdots,\alpha_k|
\delta_1,\cdots,\delta_p|\beta_1,\cdots,\beta_n|
\gamma_1,\cdots,\gamma_r)\\
&=&
\frac{
\langle B_\epsilon| 
\widetilde{V}'(\alpha_k)
\cdots \widetilde{V}'(\alpha_1)
\bar{V}'(\delta_p)\cdots
\bar{V}'(\delta_1)
V(\beta_n)\cdots V(\beta_1)
\bar{V}(\gamma_r)\cdots \bar{V}(\gamma_1)
|B_\epsilon\rangle }
{\langle B_\epsilon | B_\epsilon \rangle }.\nonumber
\end{eqnarray}
Taking the normal ordering, we have
\begin{eqnarray}
&&R_\epsilon(\alpha_1,\cdots,\alpha_k|
\delta_1,\cdots,\delta_p|\beta_1,\cdots,\beta_n|
\gamma_1,\cdots,\gamma_r)
\nonumber
\\
&=&
\delta_{2k-2p+n-2r,0}
\prod_{1\leqq j<l \leqq n}g_\epsilon(\beta_j-\beta_l)
\prod_{j=1}^k \xi_j^{2p+n-2r-1}
\prod_{j=1}^p v_j^{-n+2r-1}
\prod_{j=1}^r w_j^{-1}
\prod_{j=1}^n \zeta_j^{r-\frac{1}{4}}
\nonumber\\
&\times&\prod_{1\leqq j<l \leqq k}(1-x^2\xi_l/\xi_j)
(1-\xi_l/\xi_j)
\prod_{1\leqq j<l \leqq p}(1-x^2v_l/v_j)
(1-v_l/v_j)
\prod_{1\leqq j<l \leqq r}(1-w_l/w_j)(1-w_l/(x^2w_j))
\nonumber
\\
&\times&
\frac
{\displaystyle
\prod_{j=1}^p\prod_{l=1}^k
(1-x^2\xi_l/v_j)(1-\xi_l/v_j)
\prod_{j=1}^r\prod_{l=1}^k(1-x\xi_l/w_j)(1-\xi_l/(xw_j))
\prod_{j=1}^n\prod_{l=1}^p(1-v_l/\zeta_j)}
{\displaystyle
\prod_{j=1}^n\prod_{l=1}^k(1-\xi_l/\zeta_j)
\prod_{j=1}^r\prod_{l=1}^p(1-xv_l/w_j)(1-v_l/(xw_j))
\prod_{j=1}^r\prod_{l=1}^n(1-\zeta_l/(xw_j))
}
\nonumber\\
&\times&
I_\epsilon(
\alpha_1,\cdots,\alpha_k|
\delta_1,\cdots,\delta_p|\beta_1,\cdots,\beta_n|
\gamma_1,\cdots,\gamma_r). 
\end{eqnarray}
Here $I_\epsilon$ is defined as 
\begin{equation}
\begin{array}{cl}
&I_\epsilon(\alpha_1,\cdots,\alpha_k|
\delta_1,\cdots,\delta_p|\beta_1,\cdots,\beta_n|
\gamma_1,\cdots,\gamma_r
)\\
=&\displaystyle
\frac{\langle 0|e^{G_\epsilon}
\exp\left(\sum_{m=1}^\infty a_{-m} X_m\right) 
\exp\left(-\sum_{m=1}^\infty
a_m Y_m\right)e^{B_\epsilon}|0\rangle}{
\langle 0|e^{G_\epsilon} e^{B_\epsilon}|0\rangle}, 
\end{array}
\label{vacuum-I}
\end{equation}
where 
\begin{equation}
\begin{array}{rcl}
X_m&=&\displaystyle\frac{2}{1-x^{2m}}
\left(x^{2m}\sum_{j=1}^k \xi_j^{-m}
-x^{2m}\sum_{j=1}^p v_j^{-m}
+x^m \sum_{j=1}^r w_j^{-m}\right)
+\frac{2}{1-x^{4m}}\left(
-x^{2m}\sum_{j=1}^n \zeta_j^{-m}\right),\\
Y_m&=&\displaystyle\frac{2}{1-x^{2m}}
\left(x^{2m}\sum_{j=1}^k \xi_j^{m}
-x^{2m}\sum_{j=1}^p v_j^{m}
+x^m \sum_{j=1}^r w_j^{m}\right)
+\frac{2}{1-x^{4m}}\left(
-x^{2m}\sum_{j=1}^n \zeta_j^{m}\right).
\end{array}
\end{equation}
The vacuum expectation value (\ref{vacuum-I}) becomes
\begin{equation}
\begin{array}{cl}
&I_\epsilon(\alpha_1,\cdots,\alpha_k|
\delta_1,\cdots,\delta_p|\beta_1,\cdots,\beta_n|
\gamma_1,\cdots,\gamma_l)\\
=&\displaystyle
\exp\left(
\sum_{m=1}^\infty
[a_m,a_{-m}]\frac{1}{1-\alpha_m\gamma_m}
\left\{\frac{1}{2}\gamma_m X_m^2
+\frac{1}{2}\alpha_m Y_m^2
-\alpha_m \gamma_m X_m Y_m\right\}\right.\\
&\left.+\displaystyle\sum_{m=1}^\infty
\frac{1}{1-\alpha_m \gamma_m}\left\{
(\delta_m+\gamma_m\beta_m)X_m
-(\beta_m+\alpha_m\delta_m)Y_m
\right\}\right),\label{vacuum-II}
\end{array}
\end{equation}
In order to derive the formulae of the vacuum expectation value
(\ref{vacuum-II}) we prepare the coherent states,
\begin{eqnarray}
|\eta\rangle=\exp\left(
\sum_{m=1}^\infty
\frac{\eta_m a_{-m}}{[a_m,a_{-m}]}\right)
|0\rangle,~~~
\langle \bar{\eta}|=\langle 0 |
\exp\left(\sum_{m=1}^\infty
\frac{\bar{\eta}_ma_m}{[a_m,a_{-m}]}\right),
\end{eqnarray}
which enjoy the following properties: 
\begin{eqnarray}
a_n|\eta \rangle=\eta_n |\eta \rangle,~~~
\langle \bar{\eta}|a_{-n}=\bar{\eta}_n\langle\bar{\eta}|.
\end{eqnarray}
The completeness relation 
\begin{eqnarray}
id=\int \prod_{m>0}
\frac{d\eta_m d\bar{\eta}_m}{[a_m,a_{-m}]}
\exp\left(
-\sum_{m=1}^\infty
\frac{\eta_m \bar{\eta}_m}{[a_m,a_{-m}]}\right)
|\eta\rangle \langle \bar{\eta}|,
\end{eqnarray}
can be easily proved. Here the integral $\int d\eta d\bar{\eta}$
implies $\displaystyle\frac{1}{2i}
\int_{-\infty}^{\infty}\int_{-\infty}^{\infty}dx dy$. 

By using the Bogoliubov transformation, we obtain
\begin{equation}
\begin{array}{cl}
&\displaystyle\langle 0 |e^{G_\epsilon}\exp\left(
\sum_{m=1}^\infty a_{-m}X_m\right)
\exp\left(-\sum_{m=1}^\infty a_m Y_m
\right) e^{B_\epsilon}|0\rangle
\\
=&\displaystyle
\langle 0 |e^{G_\epsilon}e^{B_\epsilon}\exp\left(
\sum_{m=1}^\infty a_{-m}\left(X_m
-\alpha_m Y_m\right)\right)
|0\rangle \\
\times &\displaystyle\exp\left(-\sum_{m=1}^\infty
\beta_m Y_m\right)
\exp\left(\frac{1}{2}\sum_{m=1}^\infty
[a_m,a_{-m}]\alpha_m Y_m^2\right).
\end{array}
\end{equation}
Inserting the completeness relation between
$e^{G_\epsilon}$ and $e^{B_\epsilon}$ and
calculating the Gaussian-type integral,
we have (\ref{vacuum-II}), and
\begin{eqnarray}
\langle B_\epsilon |
B_\epsilon \rangle
=
\prod_{m=1}^\infty
\frac{1}{\sqrt{1-\alpha_m \gamma_m}}
\exp\left(\frac{1}{2}\sum_{m=1}^\infty
\frac{1}{1-\alpha_m \gamma_m}
\frac{1}{[a_m,a_{-m}]}\left(
\alpha_m \delta_m^2+\gamma_m \beta_m^2
+2\beta_m \delta_m\right)\right)
\end{eqnarray}

The vacuum expectation value (\ref{vacuum-II})
is evaluated by 
\begin{eqnarray}
\exp\left(-\sum_{m=1}^\infty
\frac{1}{m}\frac{z^m}{(1-p_1^m)\cdots (1-p_n^m)}\right)
&=&(z;p_1, \cdots , p_n)_\infty,~~~~ |p_i|<1\,
~~(1\leqq i\leqq n).
\end{eqnarray}
The norm of the vacuums is given as 
\begin{equation}
\begin{array}{rcl}
\langle B_\epsilon |
B_\epsilon \rangle
=
\displaystyle\frac{(x^2r^{-2};x^4)_\infty 
(x^{10}r^{-2};x^8)_\infty}
{(x^6;x^8)_\infty (x^8r^{-2};x^8)_\infty}.
\end{array}
\end{equation}
The integral $I_\epsilon(\alpha_1,\cdots,\alpha_k|
\delta_1,\cdots,\delta_p|
\beta_1,\cdots,\beta_n|
\gamma_1,\cdots,\gamma_r)$ is given by 
\begin{equation}
\begin{array}{cl}
&I_\epsilon(\alpha_1,\cdots,\alpha_k|
\delta_1,\cdots,\delta_p|
\beta_1,\cdots,\beta_n|
\gamma_1,\cdots,\gamma_r)\\
=&\displaystyle
(x^4;x^2)_\infty^{k+p}
(x^2;x^2)_\infty^r
\left(\frac{(x^4;x^4,x^4)_\infty}
{(x^6;x^4,x^4)_\infty}\right)^n
\\
\times&\displaystyle
\prod_{j=1}^k
(x^2\xi_j^2;x^4)_\infty
(x^6\xi_j^{-2};x^4)_\infty
\prod_{j=1}^p
(v_j^2;x^4)_\infty(x^4v_j^{-2};x^4)_\infty 
\prod_{j=1}^r
(w_j^2;x^4)_\infty (x^4w_j^{-2};x^4)_\infty
\\
\times&\displaystyle
\prod_{j=1}^n
\sqrt{\frac{(\zeta_j^2;x^4,x^4)_\infty
(x^4\zeta_j^{-2};x^4,x^4)_\infty (\zeta_j^2;x^8)_\infty
(x^4\zeta_j^{-2};x^8)_\infty}
{(x^2\zeta_j^2;x^4,x^4)_\infty
(x^6\zeta_j^{-2};x^4,x^4)_\infty
(x^2\zeta_j^2;x^8)_\infty (x^6\zeta_j^{-2};x^8)_\infty}}
\\
\times&\displaystyle
\prod_{j=1}^k(1-r^{-1}\xi_j)
\prod_{j=1}^p\frac{1}{1-r^{-1}v_j}
\prod_{j=1}^r(1-r^{-1}x^{-1}w_j)
\prod_{j=1}^n \frac{(x^2r^{-1}\zeta_j;x^4)_\infty}
{(r^{-1}\zeta_j;x^4)_\infty}\\
\times&\displaystyle
\prod_{1\leqq i<j \leqq k}
(\xi_i \xi_j;x^2)_\infty
(x^4\xi_i^{-1} \xi_j^{-1};x^2)_\infty
(x^4\xi_i \xi_j^{-1};x^2)_\infty
(x^4\xi_i^{-1}\xi_j;x^2)\\
\times&\displaystyle
\prod_{1\leqq i<j \leqq p}
(v_i v_j;x^2)_\infty
(x^4v_i^{-1} v_j^{-1};x^2)_\infty
(x^4v_i v_j^{-1};x^2)_\infty
(x^4v_i^{-1}v_j;x^2)\\
\times&\displaystyle
\prod_{1\leqq i<j \leqq r}
(x^{-2}w_i w_j;x^2)_\infty
(x^2w_i^{-1} w_j^{-1};x^2)_\infty
(x^2w_i w_j^{-1};x^2)_\infty
(x^2w_i^{-1}w_j;x^2)\\
\times&\displaystyle
\prod_{1\leqq i<j \leqq n}
\frac{(\zeta_i\zeta_j;x^4,x^4)_\infty
(x^4\zeta_i^{-1}\zeta_j^{-1};x^4,x^4)_\infty
(x^4\zeta_i^{-1}\zeta_j;x^4,x^4)_\infty
(x^4\zeta_i\zeta_j^{-1};x^4,x^4)_\infty
}{
(x^2\zeta_i\zeta_j;x^4,x^4)_\infty
(x^6\zeta_i^{-1}\zeta_j^{-1};x^4,x^4)_\infty
(x^6\zeta_i^{-1}\zeta_j;x^4,x^4)_\infty
(x^6\zeta_i\zeta_j^{-1};x^4,x^4)_\infty
}\\
\times&\displaystyle
\prod_{i=1}^k\prod_{j=1}^p
\left\{( \xi_i v_j;x^2)
(x^4\xi_i^{-1}v_j^{-1};x^2)_\infty
(x^4\xi_i^{-1} v_j;x^2)_\infty
(x^4\xi_i v_j^{-1};x^2)_\infty\right\}^{-1}
\\
\times&\displaystyle
\prod_{i=1}^k\prod_{j=1}^r
(x^{-1}\xi_i w_j;x^2)
(x^3\xi_i^{-1}w_j^{-1};x^2)_\infty
(x^3\xi_i^{-1} w_j;x^2)_\infty
(x^3\xi_i w_j^{-1};x^2)_\infty
\\
\times&\displaystyle\prod_{i=1}^p\prod_{j=1}^r
\left\{(x^{-1}v_i w_j;x^2)
(x^3v_i^{-1}w_j^{-1};x^2)_\infty
(x^3v_i^{-1} w_j;x^2)_\infty
(x^3v_i w_j^{-1};x^2)_\infty\right\}^{-1}
\\
\times&\displaystyle
\prod_{i=1}^k\prod_{j=1}^n
\left\{(\xi_i \zeta_j;x^4)
(x^4\xi_i^{-1}\zeta_j^{-1};x^4)_\infty
(x^4\xi_i^{-1} \zeta_j;x^4)_\infty
(x^4xi_i \zeta_j^{-1};x^4)_\infty\right\}^{-1}
\\
\times&\displaystyle
\prod_{i=1}^p\prod_{j=1}^n
(v_i \zeta_j;x^4)
(x^4v_i^{-1}\zeta_j^{-1};x^4)_\infty
(x^4v_i^{-1} \zeta_j;x^4)_\infty
(x^4v_i \zeta_j^{-1};x^4)_\infty
\\
\times&\displaystyle
\prod_{i=1}^r\prod_{j=1}^n
\left\{(x^{-1}w_i \zeta_j;x^4)
(x^3w_i^{-1}\zeta_j^{-1};x^4)_\infty
(x^3w_i^{-1} \zeta_j;x^4)_\infty
(x^3w_i \zeta_j^{-1};x^4)_\infty\right\}^{-1}.
\label{eq:I_e}
\end{array}
\end{equation}

Here we have obtained the integral representation
of the form factor in the ultraviolet regularization 
scheme. Next we consider the original problem by 
taking the limit $\epsilon \rightarrow 0$
of (\ref{eq:int-ff}). 

Let us set the function
\begin{equation}
Q_\epsilon(\beta ):=
\frac{\gamma (\zeta^2)}{(x; x^4)_\infty}, 
\end{equation}
where
$$
\gamma (\zeta^2 )=
\frac{(\zeta^{2} ; x^4, x^4)_\infty
(x^4\zeta^{-2}; x^4, x^4)_\infty}
{(x^2\zeta^{2} ; x^4, x^4)_\infty
(x^6\zeta^{-2}; x^4, x^4)_\infty}. 
$$
Then as $\epsilon \rightarrow 0$ 
the function $Q_\epsilon(\beta )$ behave like 
\begin{equation}
Q_\epsilon(\beta )
\sim\exp \left( -\int_0^\infty 
\frac{dy}{y} \frac{e^{\pi y/2} 
\left({\rm sh}^2
\left( \frac{i\beta +\pi }{2} y\right)
-{\rm sh}^2 \frac{\pi y}{4}\right)}
{{\rm sh} \pi y {\rm ch}\frac{\pi y}{2}}\right)
=:Q(\beta).
\end{equation}
In the limit $\epsilon \rightarrow 0$, 
the integral (\ref{eq:I_e}) thus behaves like 
\begin{equation}
\begin{array}{rcl}
I_\epsilon &\sim& 
\displaystyle\prod_{j=1}^k \frac{1}{
\Gamma (\frac{1}{2}+\frac{\alpha_j}{\pi i})
\Gamma (\frac{3}{2}-\frac{\alpha_j}{\pi i})}
\prod_{j=1}^p \frac{1}{
\Gamma (\frac{\delta_j}{\pi i})
\Gamma (1-\frac{\delta_j}{\pi i})} \\[6mm]
&\times & 
\displaystyle\prod_{j=1}^r \frac{1}{
\Gamma (\frac{\gamma_j}{\pi i})
\Gamma (1-\frac{\gamma_j}{\pi i})} 
\prod_{j=1}^n \sqrt{\frac{\Gamma 
(\frac{1}{4}+\frac{\beta_j}{2\pi i})
\Gamma (\frac{3}{4}-\frac{\beta_j}{2\pi i})} 
{\Gamma (\frac{\beta_j}{2\pi i})
\Gamma (\frac{1}{2}-\frac{\beta_j}{2\pi i})} 
Q(\beta_j )} \\[6mm]
&\times & 
\displaystyle\prod_{j=1}^k (\alpha_j -\mu )
\prod_{j=1}^p \frac{1}{\delta_j -\mu }
\prod_{j=1}^r (\gamma_j -\mu -\frac{\pi i}{2})
\prod_{j=1}^n \frac{\Gamma 
(\frac{\beta_j -\mu}{2\pi i})}
{\Gamma (\frac{1}{2}+\frac{\beta_j -\mu}{2\pi i})} 
\\[6mm] 
&\times & 
\displaystyle\prod_{1\leqq i<j\leqq k} 
\frac{1}{\Gamma 
(\frac{\alpha_i +\alpha_j}{\pi i})
\Gamma 
(2-\frac{\alpha_i +\alpha_j}{\pi i})
\Gamma 
(2+\frac{\alpha_i -\alpha_j}{\pi i})
\Gamma 
(2-\frac{\alpha_i -\alpha_j}{\pi i})}
\\[6mm] 
&\times & 
\displaystyle\prod_{1\leqq i<j\leqq p} 
\frac{1}{\Gamma 
(\frac{\delta_i +\delta_j}{\pi i})
\Gamma 
(2-\frac{\delta_i +\delta_j}{\pi i})
\Gamma 
(2+\frac{\delta_i -\delta_j}{\pi i})
\Gamma 
(2-\frac{\delta_i -\delta_j}{\pi i})}
\\[6mm] 
&\times & 
\displaystyle\prod_{1\leqq i<j\leqq r} 
\frac{1}{\Gamma 
(-1+\frac{\gamma_i +\gamma_j}{\pi i})
\Gamma 
(1-\frac{\gamma_i +\gamma_j}{\pi i})
\Gamma 
(1+\frac{\gamma_i -\gamma_j}{\pi i})
\Gamma 
(1-\frac{\gamma_i -\gamma_j}{\pi i})}
\\[6mm] 
&\times & 
\displaystyle\prod_{1\leqq i<j\leqq n} 
\frac{\Gamma 
(\frac{\beta_i -\beta_j}{2\pi i})}
{\Gamma 
(\frac{1}{2}+\frac{\beta_i -\beta_j}{2\pi i})}
Q\left( \frac{\beta_i +\beta_j}{2} \right)
Q\left( \frac{\beta_i -\beta_j}{2} \right)
\label{def:I}
\\[6mm]
&\times & 
\displaystyle\prod_{i=1}^k \prod_{j=1}^p 
\Gamma 
\left(\frac{\alpha_i +\delta_j}{\pi i}\right)
\Gamma 
\left(2-\frac{\alpha_i +\delta_j}{\pi i}\right)
\Gamma 
\left(2+\frac{\alpha_i -\delta_j}{\pi i}\right)
\Gamma 
\left(2-\frac{\alpha_i -\delta_j}{\pi i}\right)
\\[6mm]
&\times & 
\displaystyle\prod_{i=1}^k \prod_{j=1}^r 
\frac{1}{\Gamma 
(-\frac{1}{2}+\frac{\alpha_i +\gamma_j}{\pi i})
\Gamma
(\frac{3}{2}-\frac{\alpha_i +\gamma_j}{\pi i})
\Gamma 
(\frac{3}{2}+\frac{\alpha_i -\gamma_j}{\pi i})
\Gamma 
(\frac{3}{2}-\frac{\alpha_i -\gamma_j}{\pi i})}
\\[6mm]
&\times & 
\displaystyle\prod_{i=1}^p \prod_{j=1}^r 
\Gamma 
\left(-\frac{1}{2}+\frac{\delta_i +\gamma_j}{\pi i}\right)
\Gamma 
\left(\frac{3}{2}-\frac{\delta_i +\gamma_j}{\pi i}\right)
\Gamma 
\left(\frac{3}{2}+\frac{\delta_i -\gamma_j}{\pi i}\right)
\Gamma 
\left(\frac{3}{2}-\frac{\delta_i -\gamma_j}{\pi i}\right)
\\[6mm]
&\times & 
\displaystyle\prod_{i=1}^k \prod_{j=1}^n 
\Gamma 
(\frac{\alpha_i +\beta_j}{2\pi i})
\Gamma 
(1-\frac{\alpha_i +\beta_j}{2\pi i})
\Gamma 
(1+\frac{\alpha_i -\beta_j}{2\pi i})
\Gamma 
(1-\frac{\alpha_i -\beta_j}{2\pi i})
\\[6mm]
&\times & 
\displaystyle\prod_{i=1}^p \prod_{j=1}^n 
\frac{1}{\Gamma 
(\frac{\delta_i +\beta_j}{2\pi i})
\Gamma 
(1-\frac{\delta_i +\beta_j}{2\pi i})
\Gamma 
(1+\frac{\delta_i -\beta_j}{2\pi i})
\Gamma 
(1-\frac{\delta_i -\beta_j}{2\pi i})}
\\[6mm]
&\times & 
\displaystyle\prod_{i=1}^r \prod_{j=1}^n 
\Gamma 
\left(-\frac{1}{4}+\frac{\gamma_i +\beta_j}{2\pi i}\right)
\Gamma 
\left(\frac{3}{4}-\frac{\gamma_i +\beta_j}{2\pi i}\right)
\Gamma 
\left(\frac{3}{4}+\frac{\gamma_i -\beta_j}{2\pi i}\right)
\Gamma 
\left(\frac{3}{4}-\frac{\gamma_i -\beta_j}{2\pi i}\right) \\
&=:&I(\alpha_1,\cdots,\alpha_k|
\delta_1,\cdots,\delta_p|
\beta_1,\cdots,\beta_n|
\gamma_1,\cdots,\gamma_r). 
\end{array}
\end{equation}

Note that as $\epsilon \rightarrow 0$, 
$$
P_\epsilon (\zeta , w)\longrightarrow 
P(\beta -\gamma ), ~~~~
P'_\epsilon (\xi , v)\longrightarrow 
P'(\alpha -\delta ), 
$$
where 
$$
P(\beta )=-\frac{i\beta +\frac{\pi}{2}}
{i\beta -\frac{\pi}{2}}, ~~~~
P'(\alpha )=\frac{i\alpha -\pi}{i\alpha +\pi}. 
$$

After all, 
we obtain the following integral representation 
of the form factor for the $SU(2)$ invariant 
massive Thirring model with boundary reflection: 
\begin{equation}
\begin{array}{cl}
&G^{m_1\cdots m_k}_{a_1\cdots a_n}
(\alpha_1 , \cdots , \alpha_k |\beta_1,\cdots,\beta_n) \\
=&\delta_{2k-2p+n-2r,0}
\displaystyle\prod_{a\in A_-}\int_{-\infty}
^{\infty} \frac{d\gamma_a}{2\pi} 
(P(\beta_a - \gamma_a )+x) 
\prod_{j=a+1}^n P (\beta_j - \gamma_a ) \\
\times& \displaystyle\prod_{m\in M_{-1}}
\int_{-\infty}
^{\infty} \frac{d\delta^1_m}{2\pi} 
\int_{-\infty}
^{\infty} \frac{d\delta^2_m}{2\pi} 
\xi_m v_m^1 v_m^2 \left( 
P'(\alpha_m - \delta_m^1) P'(\alpha_m - \delta_m^2) -
2P'(\alpha_m - \delta_m^1)+1 \right) \\
\times& \displaystyle\prod_{j=m+1}^k 
P' (\alpha_m - \delta_m^1) P' (\alpha_m - \delta_m^2) 
\prod_{m' \in M_0} \int_{-\infty}
^{\infty} \frac{d\delta_{m'}}{2\pi} 
\xi_{m'}v_{m'} \left( P'(\alpha_{m'}- \delta_{m'})-1 
\right) \\
\times & \displaystyle\prod_{j=m'+1}^k 
P'(\alpha_j - \delta_{m'}) 
\prod_{1\leqq j<l \leqq n}\frac{\Gamma(\frac{1}{2}-
\frac{\beta_j-\beta_l}{2\pi i})}{\Gamma(-
\frac{\beta_j-\beta_l}{2\pi i})}
\prod_{1\leqq j<l 
\leqq k}(\pi-i(\alpha_l -\alpha_j ))
i(\alpha_j -\alpha_l )\\
\times&\displaystyle\prod_
{1\leqq j<l \leqq p}(\pi -i(\delta_l 
-\delta_j))i(\delta_j -\delta_l)
\prod_{1\leqq j<l \leqq r}i(\gamma_l -
\gamma_j)(\pi+i(\gamma_l -\gamma_j))
\\
\times&
\frac
{\displaystyle
\prod_{j=1}^p\prod_{l=1}^k
(\pi-i(\alpha_l -\delta_j))i(\delta_j -\alpha_l )
\prod_{j=1}^r\prod_{l=1}^k 
(\pi -i(\alpha_l -\delta_j))i(\delta_j -\alpha_l )
\prod_{j=1}^n\prod_{l=1}^p i(\beta_j -\delta_l )}
{\displaystyle
\prod_{j=1}^n\prod_{l=1}^k 
(\beta_j -\alpha_l )
\prod_{j=1}^r\prod_{l=1}^p 
(\frac{\pi}{2}-i(\delta_l -\gamma_j))
(-\frac{\pi}{2}-i(\delta_l -\gamma_j))
\prod_{j=1}^r\prod_{l=1}^n
(-\frac{\pi}{2}-i(\beta_l-\gamma_j))
} \\
\times&
I(\alpha_1,\cdots,\alpha_k|
\delta_1,\cdots,\delta_p|\beta_1,\cdots,\beta_n|
\gamma_1,\cdots,\gamma_r), 
\label{eq:fin}
\end{array}
\end{equation}
where $I(\alpha_1,\cdots,\alpha_k|
\delta_1,\cdots,\delta_p|\beta_1,\cdots,\beta_n|
\gamma_1,\cdots,\gamma_r)$ is 
defined in (\ref{def:I}).

In this paper we have obtained the boundary state 
and its dual of the boundary $SU(2)$ invariant 
massive Thirring model, by solving the boundary 
crossing symmetry condition. We have also presented 
the integral formulae of the soliton form factor of the model. 
For an application of our integral formulae,
the asymptotic behaviors of the soliton form factors
may be evaluated along the line of Smirnov's pioneering
work
\cite{Smbk}. 

It is known that the $SU(2)$ invariant 
massive Thirring model can be obtained from 
the sine-Gordon model by taking the limit 
such that an appropriate parameter of the 
latter model goes to infinity \cite{Smbk}. 
Nevertheless, it is not easy to see that our 
expression (\ref{eq:fin}) is reduced from the 
corresponding form factor obtained in \cite{HSWY}. 
A part of technical reasons of the difficulty is 
as follows. For the boundary sine-Gordon model 
the integral contour for the asymptotic generators 
of the ZF algebra 
is more complicated than that for the local generators. 
On the other hand, for the present case, the integral 
contours for both the asymptotic and local generators 
of the ZF algebra can be taken on the real line. 

In this paper we used the diagonal solution of the 
boundary Yang--Baxter equation. 
It may be interesting to 
study a 
nondiagonal boundary condition case.

~\\
{\it Acknowledgements}~~
TK was partly supported by Grant-in-Aid for
Encouragements for Young Scientists (A)
from Japan Society for the Promotion of Science.
(11740099)

\end{document}